\documentclass[conference]{ieeeconf}%{IEEEtran}
\IEEEoverridecommandlockouts
\usepackage{cite}
\usepackage{amsmath,amssymb,amsfonts}
\usepackage{algorithmic}
\usepackage{lipsum}
\usepackage{multirow}
\usepackage{graphicx}
\usepackage[caption=false]{subfig}
\usepackage{textcomp}
\usepackage{xcolor}
\def\BibTeX{{\rm B\kern-.05em{\sc i\kern-.025em b}\kern-.08em
    T\kern-.1667em\lower.7ex\hbox{E}\kern-.125emX}}
\begin{document}

\title{\LARGE \bf Human-vehicle interaction for autonomous vehicles in crosswalk scenarios: Field experiments with pedestrians and passengers}

%Communicative Interfaces for Autonomous Vehicles in Crosswalk interactions. Pedestrian and Passenger User Acceptance Study}

% Human-vehicle interfaces for autonomous vehicles in crosswalk interactions: Field experiments with pedestrians and passengers.

% Autonomous vehicle communication with pedestrians and passengers in crosswalk scenarios: field experiments.

%Communication interfaces for autonomous vehicles in Crosswalk interactions: Pedestrian and Passenger Field Experiments

% Human-Vehicle Interfaces for Autonomous Vehicles in Crosswalk interactions: Pedestrian and Passenger Field Experiments

\author{R. Izquierdo$^{1}$, S. Martín$^{1}$,  J. Alonso$^{1}$, I. Parra$^1$, M. A. Sotelo$^1$, D. Fern\'andez-Llorca$^{1,2}$
\thanks{$^{1}$Computer Engineering Department, Universidad de Alcal\'a, Alcal\'a de Henares, Spain.
        {\tt\small ruben.izquierdo@uah.es}}% 
        \newline
\thanks{$^{2}$European Commission, Joint Research Centre, Seville, Spain.}
}

% \author{
% \IEEEauthorblockN{Rubén Izquierdo}
% \IEEEauthorblockA{\textit{Computer Engineering dept.} \\
% \textit{Universidad de Alcalá}\\
% ruben.izquierdo@uah.es}
% \and
% \IEEEauthorblockN{Sergio Martín Serrrano}
% \IEEEauthorblockA{\textit{Computer Engineering dept.} \\
% \textit{Universidad de Alcalá}}
% \and
% \IEEEauthorblockN{David Fernández-Llorca}
% \IEEEauthorblockA{{Joint Research Center} \\
% \textit{Universidad de Alcalá}}
% \and
% \IEEEauthorblockN{Miguel Ángel Sotelo}
% \IEEEauthorblockA{\textit{Computer Engineering dept.} \\
% \textit{Universidad de Alcalá}}
% }

\maketitle

%%%%%%%%%%%%%%%%%%%%%%%%%%%%%%%%%%%%%%%%%%%%%%%%%%%%%%%%%%%%%%%%%%%%%%%%%%%%%%%
%%%%%%%%%%%%%%%%%%%%%%%%%%%%%%%%%%%%%%%%%%%%%%%%%%%%%%%%%%%%%%%%%%%%%%%%%%%%%%%

\textbf{This work has been accepted for publication at the 26\textsuperscript{nth} IEEE International Conference on Intelligent Transportation Systems.}\\
\begin{abstract}
This paper presents the results of real-world testing of human-vehicle interactions with an autonomous vehicle equipped with internal and external Human Machine Interfaces (HMIs) in a crosswalk scenario. The internal and external HMIs were combined with implicit communication techniques using gentle and aggressive braking maneuvers in the crosswalk. 
Results have been collected in the form of questionnaires and measurable variables such as distance or speed when the pedestrian decides to cross.
The questionnaires show that pedestrians feel safer when the external HMI or the gentle braking maneuver is used interchangeably, while the measured variables show that external HMI only helps in combination with the gentle braking maneuver.
The questionnaires also show that internal HMI only improves passenger confidence in combination with the aggressive braking maneuver.
\end{abstract}

%%%%%%%%%%%%%%%%%%%%%%%%%%%%%%%%%%%%%%%%%%%%%%%%%%%%%%%%%%%%%%%%%%%%%%%%%%%%%%%
%%%%%%%%%%%%%%%%%%%%%%%%%%%%%%%%%%%%%%%%%%%%%%%%%%%%%%%%%%%%%%%%%%%%%%%%%%%%%%%

\section{Introduction}
Efficient human-vehicle interaction in the context of Autonomous Vehicles (AVs) 
%\footnote{We use Autonomous Vehicle to refer to Highly Automated or Driverless vehicles, where users inside the vehicle are mere passengers.} 
has a fundamental impact on the user's sense of agency, perception of risk, and trust \cite{Li2019}. These factors, in turn, are essential to avoid both disuse and misuse of technology, which directly affect user acceptance and safety respectively \cite{Llorca2021}. 

Human-vehicle interaction in autonomous driving is a multi-user problem which, at a first level, includes mainly two groups of humans: those using the AV (i.e., passengers) and external road users interacting with the AV (i.e., pedestrians, cyclists, drivers). The lack of a driver to communicate with, both from the point of view of a passenger \cite{Detjen2021} and an external road agent \cite{Rasouli2020}, changes the nature and dynamics of interactions. In this new context, AVs need to communicate their intentions to non-automated or non-connected elements using all available resources. This communication is particularly important in scenarios where safety-relevant interactions occur, such as scenarios with pedestrians crossing in front of the AV.

\begin{figure}
\centerline{\includegraphics[width=\columnwidth]{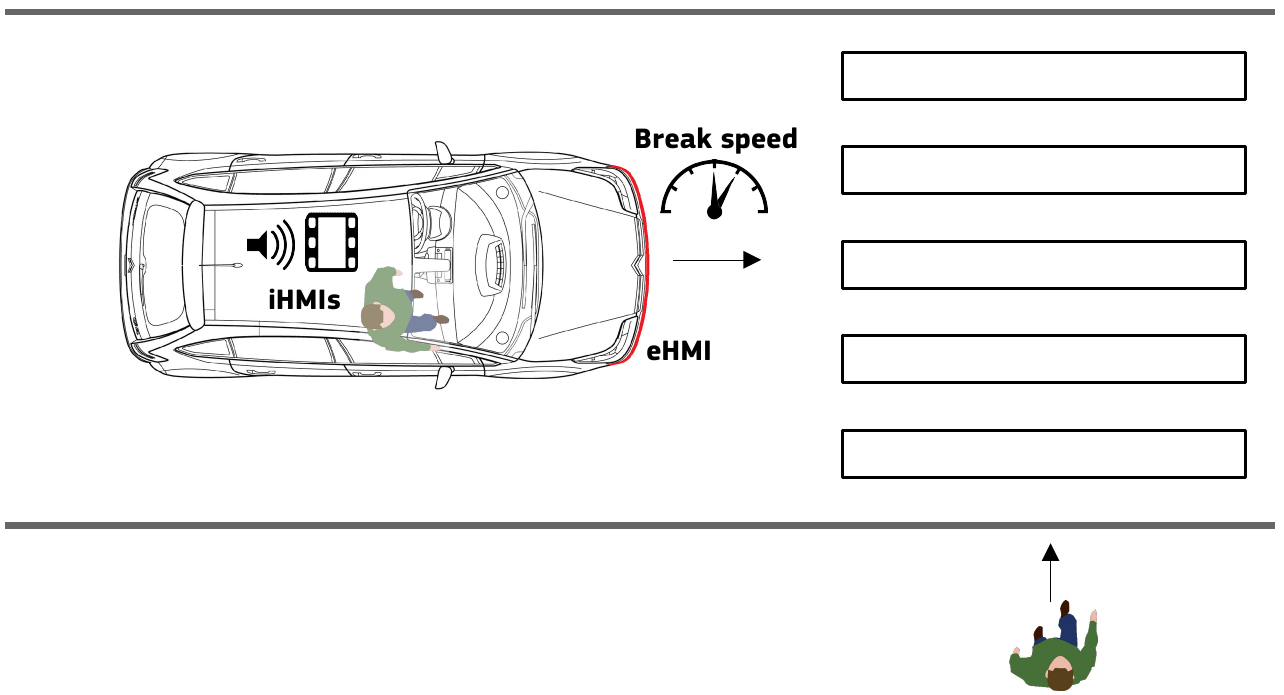}}
\caption{Experiment configuration. The vehicle drives autonomously with a passenger towards the crosswalk while the pedestrian walks perpendicularly to the road and crosses.}
\label{fig:schematic1}
\vspace{-0.42cm}
\end{figure}

The use of Vehicle-to-everything (V2X) \cite{ignacio_v2v} endows communications between other automated agents, such as other connected vehicles and the infrastructure, but still keeps humans unaware of the vehicle’s intentions. Human-vehicle interaction is mainly developed through human-machine interfaces (HMIs), both internal (iHMI) and external (eHMI), whose specific modality is linked to vehicle technology and human capabilities \cite{Llorca2021}. The behavior of the vehicle, i.e. its movement dynamics, is also an important form of implicit communication with a considerable impact on the interaction \cite{Rasouli2020}, \cite{Dey2021}. 

The impact of these forms of explicit or implicit communication on in-vehicle users (drivers or passengers) and other external road users, although widely studied, has always been done separately, which does not allow holistic conclusions to be drawn. From an experimental point of view, previous work focuses on simulated environments using virtual reality \cite{Martin2023}, or on real environments with two main types of constraints. On the one hand, cases in which the pedestrian only expresses an intention to cross without actually performing the crossing action \cite{Dey2021}. On the other hand, cases where the driving is not really automatic but mediated by Wizard of Oz methods \cite{Lagstrom2015}. In all cases, the results are somehow limited due to the mismatch with real-world interaction scenarios.

In this work, we present the results of a real field study on human-vehicle interaction in crosswalk scenarios, involving both pedestrians and passengers. Our automated test vehicle \cite{drivertive2018} is not mediated by the Wizard of Oz approach. The pedestrians does not explicitly communicate their intention to cross (with the added difficulty of identifying the exact moment when the pedestrian decided to cross), but decide to cross (or not), and finalize their crossing action. In this way, on the one hand, we can draw conclusions that take into account the two types of users who interact with the AV in a holistic way. It also allows us to investigate the impact of previous experience of interaction as a passenger on pedestrian behavior and vice versa. On the other hand, we can minimize the gap between the interactions measured in our experimental setup and those that would occur in a real environment. We evaluate different types of internal and external HMIs, as well as implicit communication, and we use both behavioral and attitudinal methods.

\section{Related work}

Studies used to assess human behavior in traffic scenes are mainly based on surveys which may not accurately collect the desired information\cite{dijksterhuis2015impact}. Recording-based studies allow data to be obtained by direct measurement and allow certain biases to be eliminated \cite{tom2011gender}. Other technologies also allow recording eye-tracking \cite{Dey2021} to analyze the focus of attention of the study subject and which factors are most relevant. In this study, both questionnaires and internal and external video recordings are used to analyze interactions.

In the eHMI section, we can find several features \cite{bazilinskyy2019survey}: anthropomorphic, textual, light patterns or trajectory projection on the ground \cite{chang2022can, modes5, mason2022lighting}. The color and shape of the communication play a fundamental role in the interaction. Messages using red color are more suitable to indicate risk, and green ones to indicate a safe situation \cite{bazilinskyy2019survey}. This idea reinforces the theories of social constructivism \cite{lynch2016social}. However, in \cite{Dey2021} a turquoise LED strip is used to convey different AV behaviors varying the light pattern but not the color. Turquoise color is commonly used as non-related-to-traffic color for "new" AV features. 

The iHMI has the advantage of being integrated inside the vehicle cabin and can communicate information to passengers through visual and/or auditory clues. In the early stages, some prototypes were developed to share situational awareness with AV researchers \cite{benderius2017best}. However for highly automated vehicles and considering people inside the cabin as mere passengers new iHMI has been developed with the goal to improve user experience, acceptance, and trust \cite{murali2022intelligent}. The iHMI has been in deep study to facilitate transferring from autonomous to manual mode and vice versa. But, to the best of our knowledge, there are no studies that have evaluated the effect of an iHMI when simultaneously interacting with other road users in a safety-critical scenario. In this work, we present results collected from real experimentation with an AV driving autonomously and interacting with a pedestrian in a crosswalk using simultaneously an iHMI and an eHMI. 

\begin{figure}
\centering
\subfloat[Screen showing no information.]{\includegraphics[width=0.5\columnwidth]{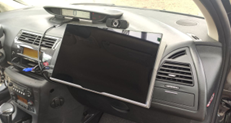}%
\label{fig:internal_off}}
\hfil
\subfloat[Image shown at manual mode.]{\includegraphics[width=0.5\columnwidth]{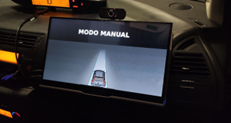}%
\label{fig:internal_manual}}
\hfil
\subfloat[Image shown at autonomous mode.]{\includegraphics[width=0.5\columnwidth]{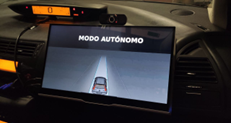}%
\label{fig:internal_auto}}
\hfil
\subfloat[Live video input with pedestrian detection.]{\includegraphics[width=0.5\columnwidth]{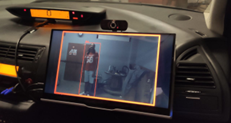}%
\label{fig:internal_ped}}
\caption{Examples of the four possible statuses of the iHMI.}
\label{fig:internal_HMI}
\end{figure}

%%%%%%%%%%%%%%%%%%%%%%%%%%%%%%%%%%%%%%%%%%%%%%%%%%%%%%%%%%%%%%%%%%%%%%%%%%%%%%%
%%%%%%%%%%%%%%%%%%%%%%%%%%%%%%%%%%%%%%%%%%%%%%%%%%%%%%%%%%%%%%%%%%%%%%%%%%%%%%%

\section{Experiment Description}
The experiment has the goal to create an interaction situation between an AV and its passenger with a pedestrian crossing through a crosswalk. The recreation of this situation under controlled  conditions allows us to measure and evaluate their behavior through direct and indirect measurements and questions. This section provides a description of the recreated situation and different experiment configurations.

\subsection{Experiment Configuration}
Figure \ref{fig:schematic1} shows a schematic representation of the experiment configuration. The pedestrian walks perpendicularly to the road with the goal to cross the road in the crosswalk. At the same time, the AV arrives at the crosswalk, and the interaction between the AV, its passenger, and the pedestrian takes place. The vehicle is driven autonomously along a straight stretch of road. The vehicle speed is 30 km/h until it starts the braking maneuver using a constant deceleration. The passenger is located in the co-drivers seat. In the driver's position, there is a backup driver to comply with current legislation but will not intervene unless critical and imminent risk. The pedestrian waits at the crosswalk with his back to the road until s/he receives the order to turn around and walk towards the pedestrian crossing. The position of the vehicle is used as a reference to trigger the pedestrian's reaction.   

\begin{figure}
\centering
\subfloat[GRAIL at solid green status.]{\includegraphics[width=0.5\columnwidth]{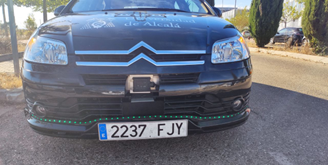}%
\label{fig:grail_green}}
\hfil
\subfloat[GRAIL at solid red status.]{\includegraphics[width=0.5\columnwidth]{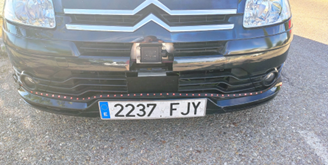}%
\label{fig:grail_red}}
\caption{Examples of the activated statuses of the eHMI.}
\label{fig:external_HMI}
\end{figure}

\subsection{Test Variations}
The autonomous vehicle has been equipped with two communication interfaces. The iHMI is a screen with the goal to communicate the current status of the vehicle to the passenger. Figure \ref{fig:internal_HMI} shows the four possible states of the iHMI. Messages shown in figures from \ref{fig:internal_manual} to \ref{fig:internal_ped} are accompanied by audio messages describing the current status of the vehicle. The messages say: "Autonomous mode activated", "Autonomous mode deactivated", and "Pedestrian detected", respectively. The detection of the pedestrian is represented with a live-video stream accompanied by the detected bounding box of the pedestrian using a state-of-the-art CNN \cite{he2017mask}.

The eHMI, so-called GRAIL (Green Assistant Interfacing Light) \cite{GRAIL} is a LED strip located in the front bumper and it is used to communicate the vehicle's intentions to other surrounding agents. Figures \ref{fig:grail_green} and \ref{fig:grail_red} show the two possible states in addition to shutdown. The GRAIL interface remains solid red when the vehicle drives at its cruise speed and turns solid green when it begins to brake and yields to pedestrians.

In addition, we introduce other implicit forms of communication such as speed profiling. The way the vehicle decreases its speed and the anticipation of the maneuver have an important role to play when interacting with pedestrians in a crosswalk. With this goal in mind, we have defined two different braking maneuvers. One is denoted as gentle or optimal with low deceleration and high anticipation. On the other hand, the aggressive or reactive braking maneuver has a higher deceleration and lower anticipation. Both braking maneuvers start at 30 km/h and end with a full stop right before the crosswalk with two deceleration values, -0.9 m/s\textsuperscript{2} for the gentle braking maneuver and -1.8 m/s\textsuperscript{2} for the aggressive one.

\subsection{Test Batch}
Five tests have been defined to evaluate the perceived level of confidence and safety using different combinations of communication techniques. Table \ref{tab:testsetup} shows the type of implicit and/or explicit ways of communication used for each test. Test number 0 is a drive-through the crosswalk area with no stop or speed reduction. The goal of this test is to prime both participants with the idea that there is a real risk and the vehicle may not stop. All tests were performed in a predefined random order except for test number 0, which was always performed in the first position.

\begin{table}[htbp]
\renewcommand{\arraystretch}{1.1}
\caption{Configuration of Experimentation Tests}
\begin{center}
\begin{tabular}{c|c|c|c|c}
%\hline
\textbf{Test}& \textbf{Braking} & \multicolumn{2}{c|}{\textbf{Explicit}} & \textbf{Stop}\\
%\cline{3-4} 
\textbf{Number} & \textbf{Maneuver} & \textbf{Internal}& \textbf{External} & \\
\hline
0   & -             & -     & -     & No \\
1   & Gentle        & -     & -     & Yes\\
2   & Aggressive    & -     & -     & Yes\\
3   & Gentle        & HMI   & GRAIL & Yes\\
4   & Aggressive    & HMI   & GRAIL & Yes\\
%\hline
\end{tabular}
\label{tab:testsetup}
\end{center}
\end{table}

\subsection{Experiment Sensing Setup}
From the experiment point of view, different sources of information are needed to measure the interaction between the AV and the participants.
The AV is equipped with an RTK-GPS system that provides precise positions of the vehicle with respect to the crosswalk. Two cameras complete the sensor setup used for the experimentation. The internal camera is mounted over the iHMI and records the passenger seat area. The external camera is mounted on the top of the vehicle and records the environment in front of the AV. The image of the pedestrian and its detection is shown in the iHMI when it is used. 

\subsection{Participants}
Volunteers were recruited at the University area, including also family and friends. A total of 34 people joined the experiment but 2 of them could not complete the whole set and their information was discarded. The sample is $N=32$ (18 men and 14 women) with an age distribution $\mu=39.7$ and $\sigma=12.6$ years.

Volunteers participated in the experiment as couples. One performed the passenger role while the other performed the pedestrian role. After finishing the complete set of tests, the participants swapped roles to perform the complete set of tests again in the same random order. This mechanism allows us to analyze whether the role initially played has any effect on the confidence perceived when interacting with the AV.

%%%%%%%%%%%%%%%%%%%%%%%%%%%%%%%%%%%%%%%%%%%%%%%%%%%%%%%%%%%%%%%%%%%%%%%%%%%%%%%
%%%%%%%%%%%%%%%%%%%%%%%%%%%%%%%%%%%%%%%%%%%%%%%%%%%%%%%%%%%%%%%%%%%%%%%%%%%%%%%

\section{Experiment Evaluation}
This section describes the tools used to assess participants' perceived confidence. For this purpose, a questionnaire was used with questions related to the perceived level of confidence, communication devices, and braking maneuvers. In addition, direct measurement variables were used during the experiments to evaluate not only what people think but also what they actually do.

\subsection{Questionnaire}
A questionnaire has been designed to ask participants about each interaction with the vehicle. As a pedestrian, three questions were answered after each experiment:
\begin{itemize}
\item[--] Q1: \emph{What was your level of confidence that the vehicle would stop and yield to you?}
%1 - No confidence, 2 - Very little confidence, 3 - Little confidence, 4 - Medium confidence, 5 - Quiet a lot of confidence, 6 - A lot of confidence, and 7 - Total confidence.
% \begin{enumerate}
% \item No confidence
% \item Very little confidence
% \item Little confidence
% \item Medium confidence
% \item Quiet a lot of confidence
% \item A lot of confidence
% \item Total confidence
% \end{enumerate}

\item[--] Q2: \emph{How did you perceive the braking of the vehicle?}
%1 - Too conservative, 2 - Quiet conservative, 3 - Somewhat conservative, 4 - Adequate, 5 - Somewhat aggressive, 6 - Quite aggressive, and 7 - Too aggressive. 
% \begin{enumerate}
% \item Too conservative
% \item Quiet conservative
% \item Somewhat conservative
% \item Adequate
% \item Somewhat aggressive
% \item Quite aggressive
% \item Too aggressive
% \end{enumerate}

\item[--] Q3: \emph{Has the visual communication interface improved your confidence to cross?}
%0 - Do not perceive any visual signal, 1 - Not at all, 2 - Very little, 3 - A little, 4 - Somewhat, 5 - Quite a lot, 6 - A lot, and 7 - Very much.
% \begin{enumerate}
% \setcounter{enumi}{-1}
% \item Do not perceive any visual signal
% \item Not at all
% \item Very little
% \item A little
% \item Somewhat
% \item Quite a lot
% \item A lot
% \item Very much 
% \end{enumerate}
\end{itemize}

As a passenger, four questions were answered: 
\begin{itemize}
\item[--] Q4: \emph{What has been your confidence in the vehicle?}
%1 - No confidence, 2 - Very little confidence, 3 - Little confidence, 4 - Medium confidence, 5 - Quiet a lot of confidence, 6 - A lot of confidence, and 7 - Total confidence.
% \begin{enumerate}
% \item No confidence
% \item Very little confidence
% \item Little confidence
% \item Medium confidence
% \item Quiet a lot of confidence
% \item A lot of confidence
% \item Total confidence
% \end{enumerate}

\item[--] Q5: \emph{How did you perceive the braking of the vehicle?}
%1 - Too conservative, 2 - Quiet conservative, 3 - Somewhat conservative, 4 - Adequate, 5 - Somewhat aggressive, 6 - Quite aggressive, and 7 - Too aggressive. 
% \begin{enumerate}
% \item Too conservative
% \item Quiet conservative
% \item Somewhat conservative
% \item Adequate
% \item Somewhat aggressive
% \item Quite aggressive
% \item Too aggressive
% \end{enumerate}

\item[--] Q6: \emph{Has the audiovisual communication interface improved the level of confidence in the vehicle?}
%0 - Do not perceive any signal, 1 - Not at all, 2 - Very little, 3 - A little, 4 - Somewhat, 5 - Quite a lot, 6 - A lot, and 7 - Very much.
% \begin{enumerate}
% \setcounter{enumi}{-1}
% \item Do not perceive any visual signal
% \item Not at all
% \item Very little
% \item A little
% \item Somewhat
% \item Quite a lot
% \item A lot
% \item Very much 
% \end{enumerate}

\item[--] Q7: \emph{Which signal was most helpful to you?}
%V - Visual, A - Audio, and B - Both.
\end{itemize}

Answers for questions from Q1 to Q6 are tabulated in a 7-step Likert scale \cite{joshi2015likert} and three answers are possible for question Q7: "visual", "audio", or "both". The answers to these questions allow us to know how participants perceived the interaction with the vehicle as passengers and as pedestrians. Note that Q3 and Q4 have a possible 0-value response in case none of the eHMI or iHMI is observed. It is intended to include a manipulation check that is answered before providing any specific value for the related question.

\subsection{Direct Measurements}
\begin{figure}
\centerline{\includegraphics[width=\columnwidth]{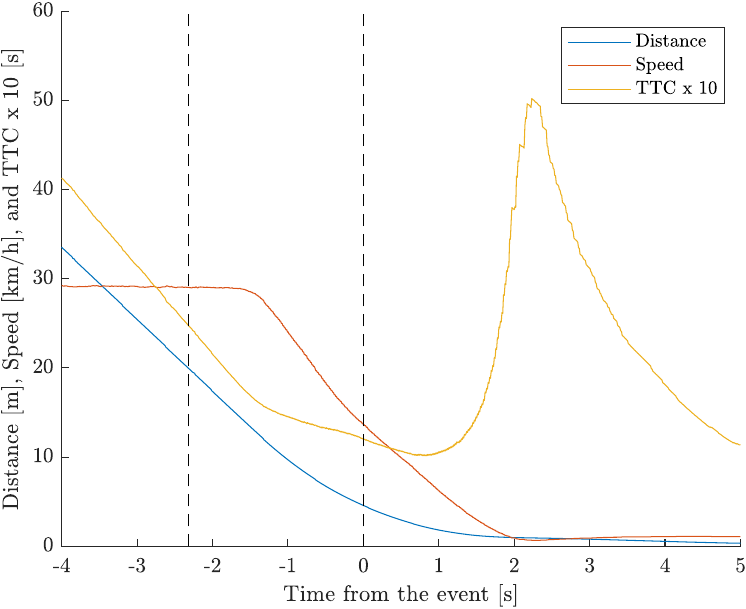}}
\caption{Example of measured variables at the crossing event.}
\label{fig:varibles_at_crossing}
\end{figure}

The distance to the pedestrian can be computed at any time during the experimentation using the vehicle positioning system. The speed of the vehicle can be directly read from the log files, and in combination with the distance, the Time To Collision (TTC) can be computed as $TTC=d/v$. These three variables can precisely describe the scene relative to the pedestrian's point of view. Figure \ref{fig:varibles_at_crossing} is an example of these three variables for an experiment with the optimal deceleration maneuver. Note that these temporal representations are meaningless without the proper labeling of the crossing event.
All trajectories are identical for each type of braking maneuver and the only thing that changes is the interaction of each participant with the vehicle, i.e. the decision to cross. 

\subsection{Crossing Decision Event}
\begin{figure}
\centerline{\includegraphics[width=\columnwidth]{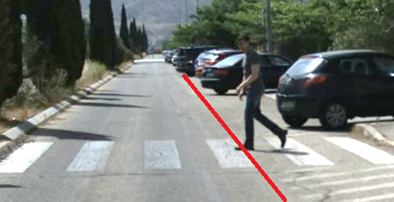}}
\caption{Crossing event example. Vehicle lane is defined by road marks unequivocally for all the experiments.}
\label{fig:vehicle_lane}
\end{figure}

\begin{figure*}
\centering
\subfloat[Initial position of pedestrian back to the crosswalk.]{\includegraphics[width=0.24\textwidth]{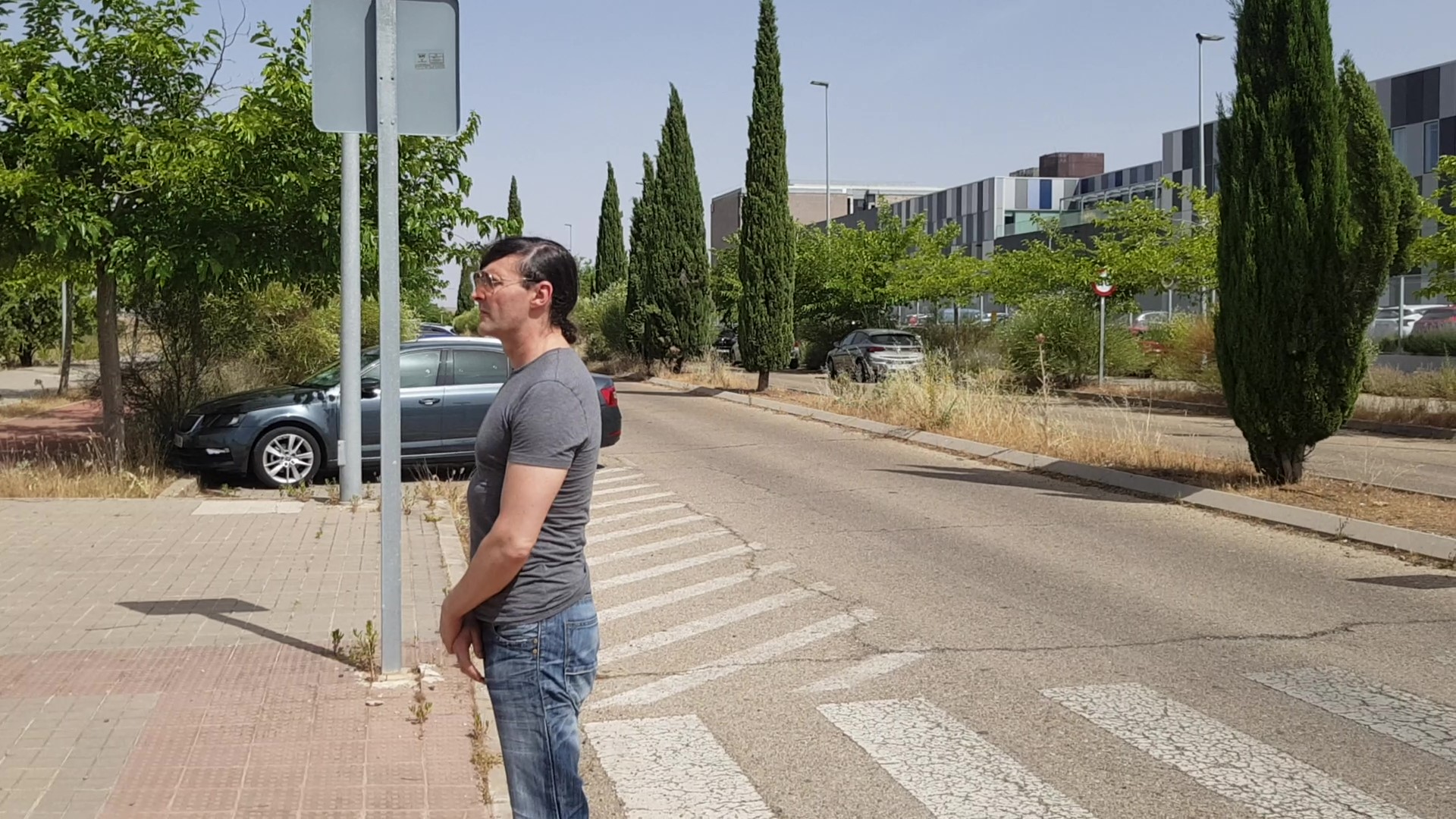}%
\label{fig_ped_1}}
\hfil
\subfloat[The pedestrian turns and faces the crosswalk.]{\includegraphics[width=0.24\textwidth]{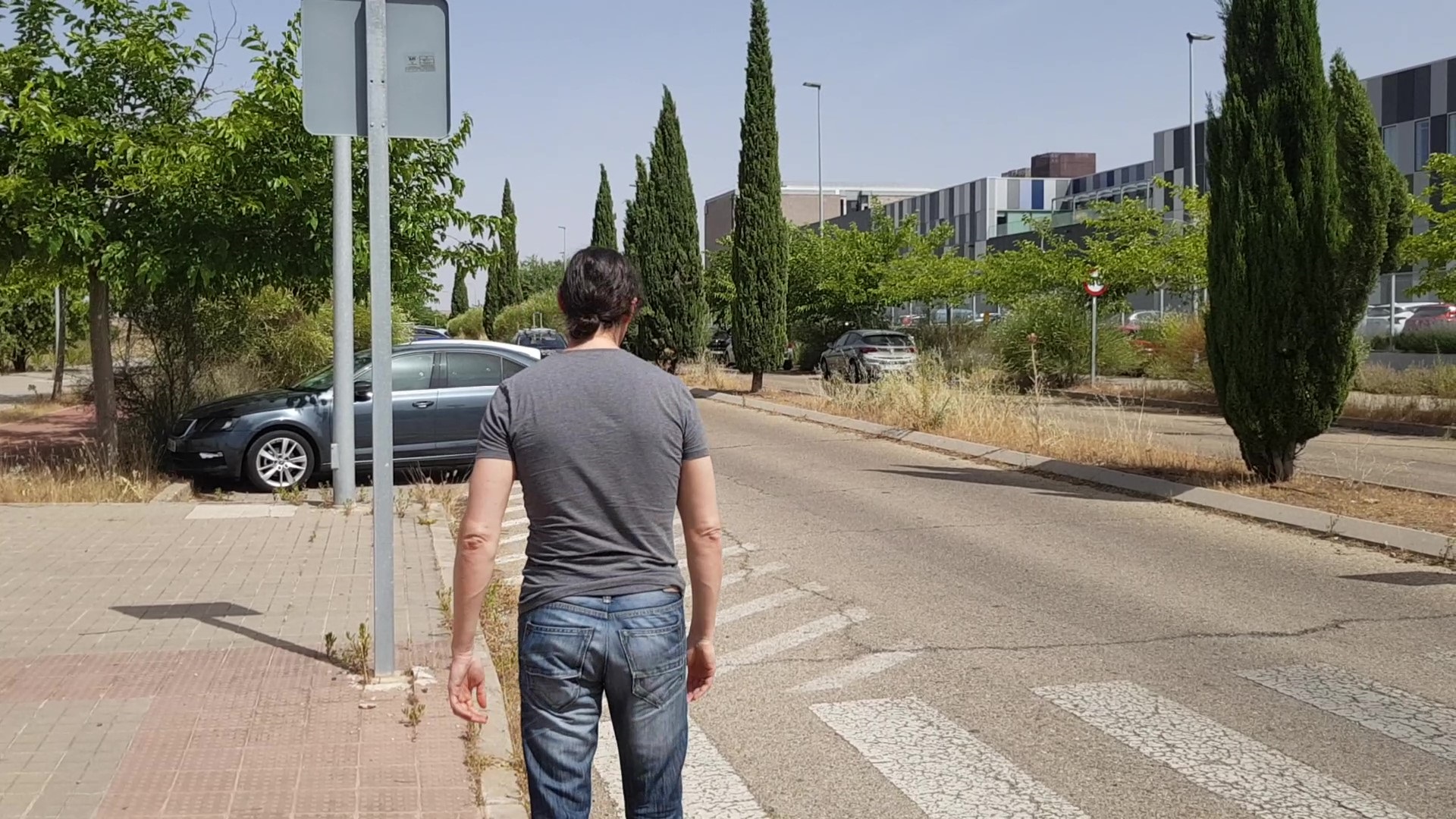}%
\label{fig_ped_2}}
\hfil
\subfloat[The pedestrian starts walking and sees the vehicle approaching.]{\includegraphics[width=0.24\textwidth]{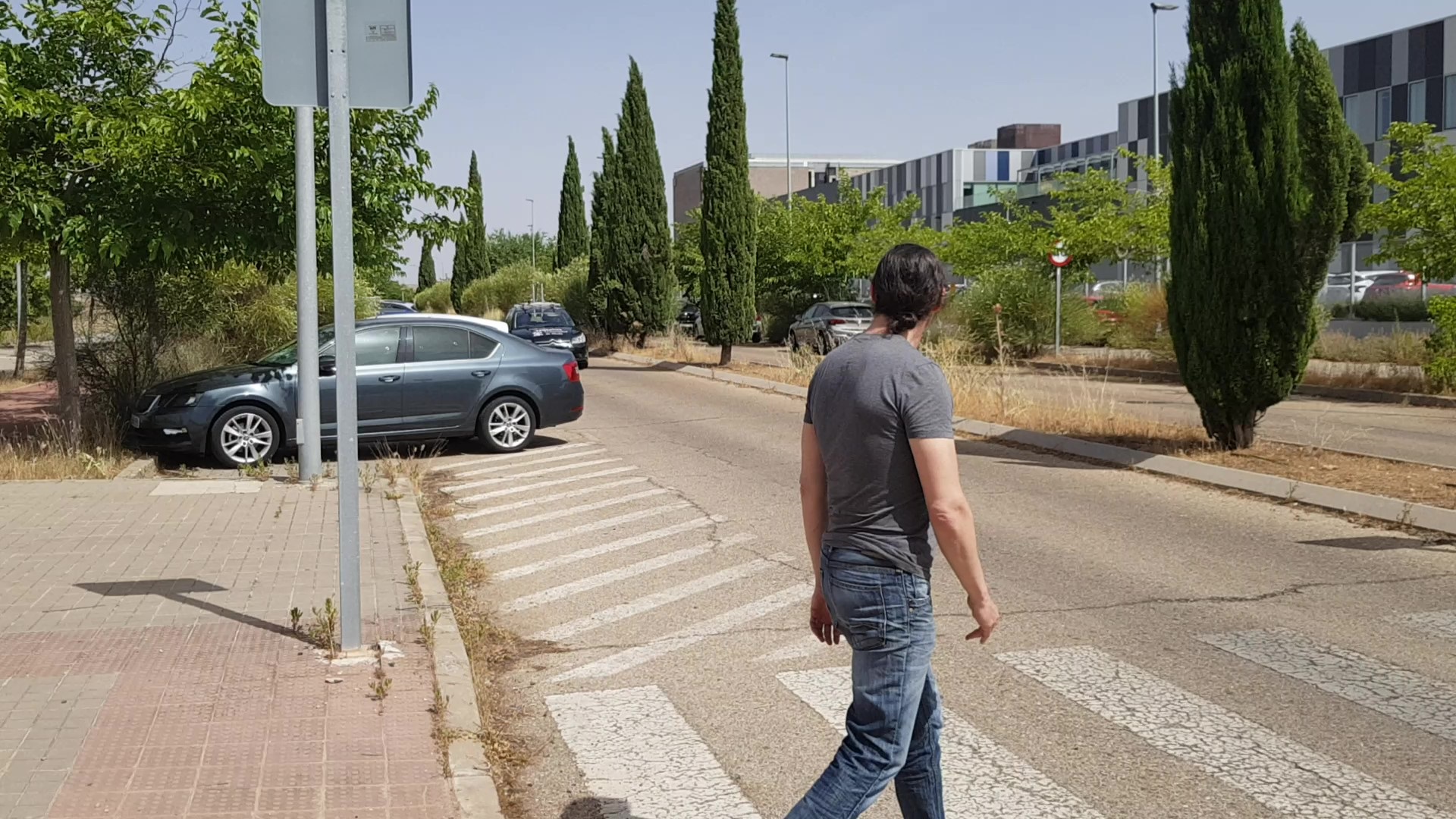}%
\label{fig_ped_3}}
\hfil
\subfloat[At this point, the pedestrian is hesitating to cross.]{\includegraphics[width=0.24\textwidth]{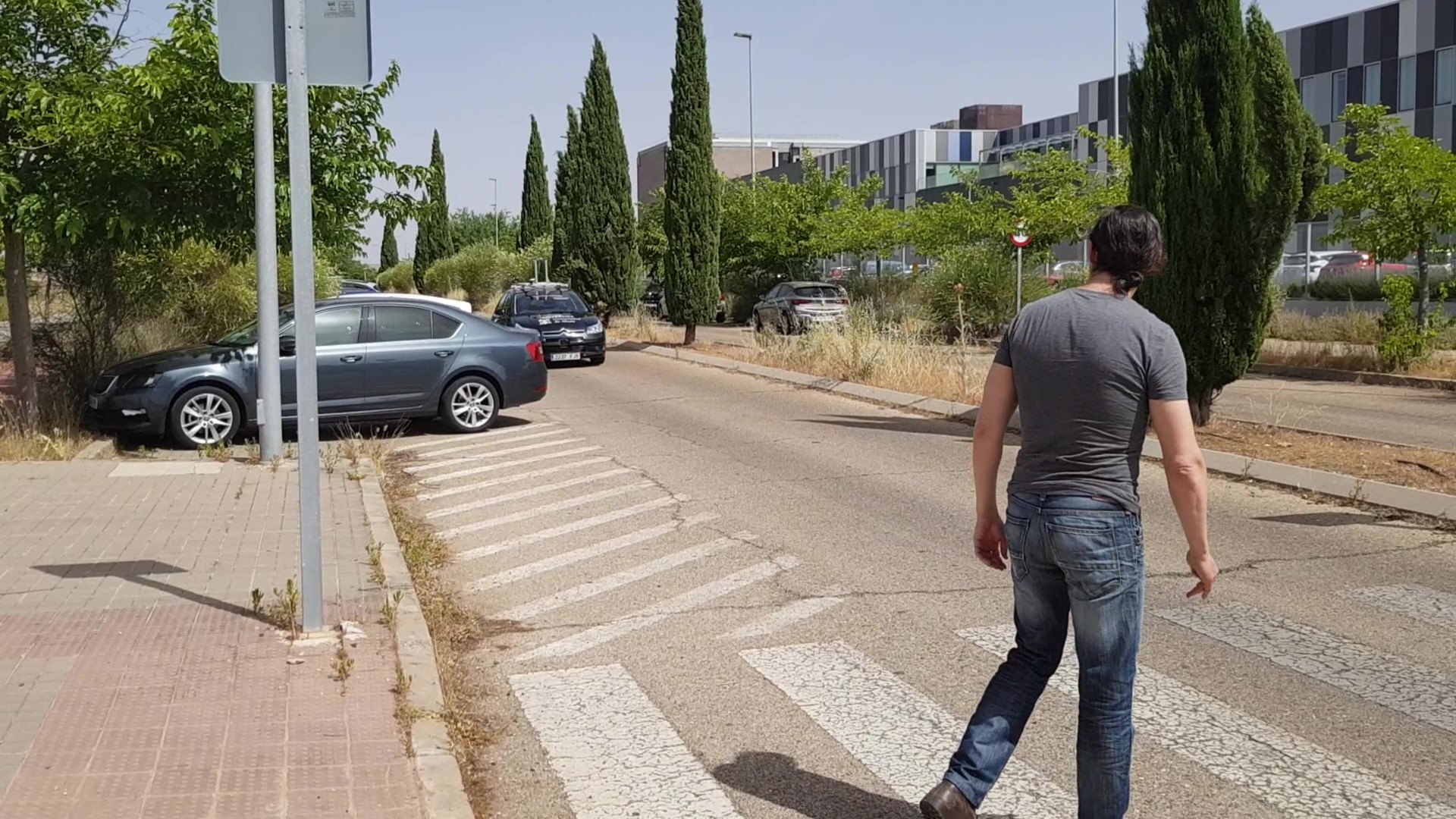}%
\label{fig_ped_4}}
\hfil
\subfloat[The pedestrian is still waiting for the vehicle's reaction.]{\includegraphics[width=0.24\textwidth]{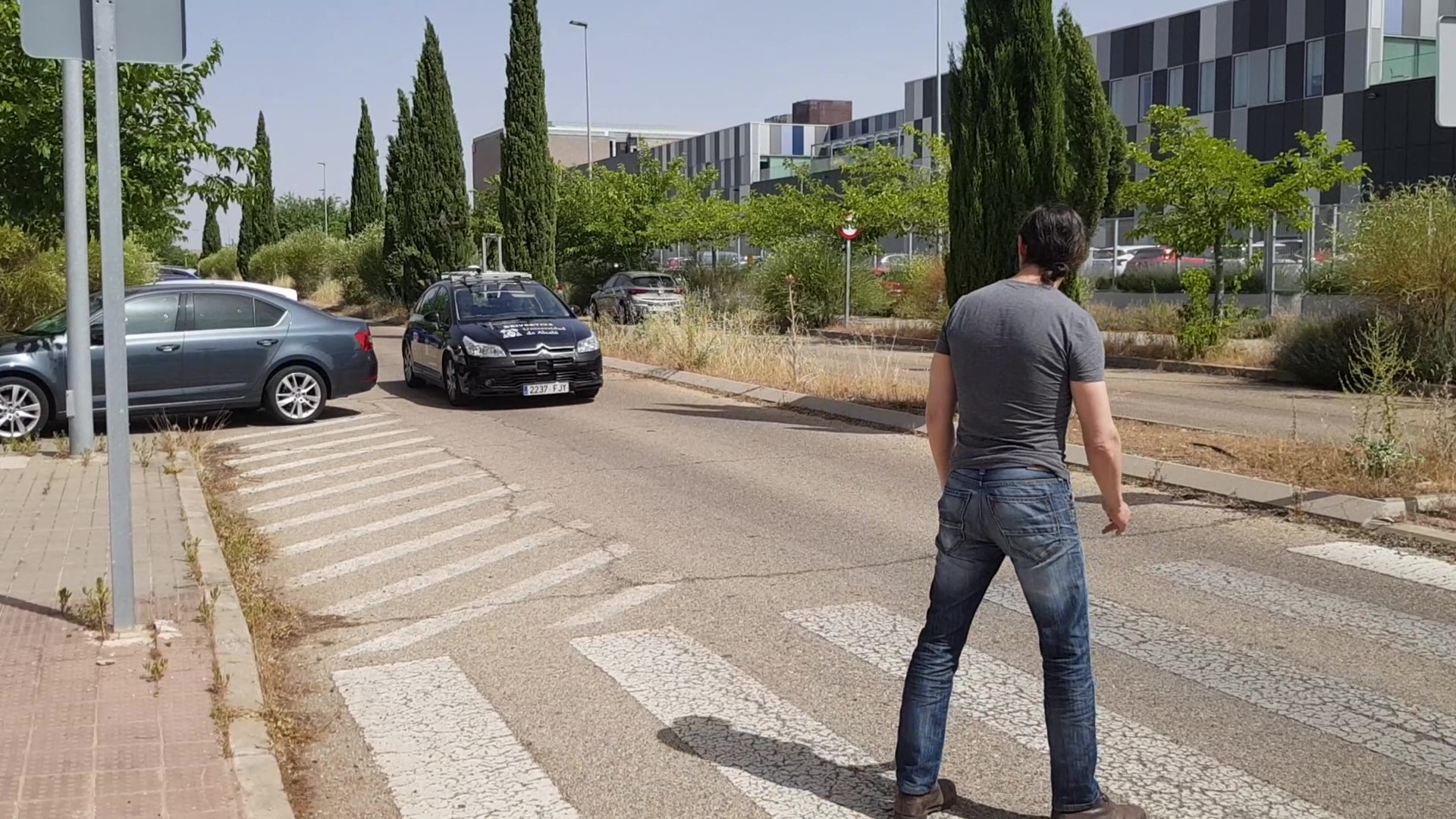}%
\label{fig_ped_5}}
\hfil
\subfloat[The pedestrian does not feel comfortable crossing while the vehicle is moving.]{\includegraphics[width=0.24\textwidth]{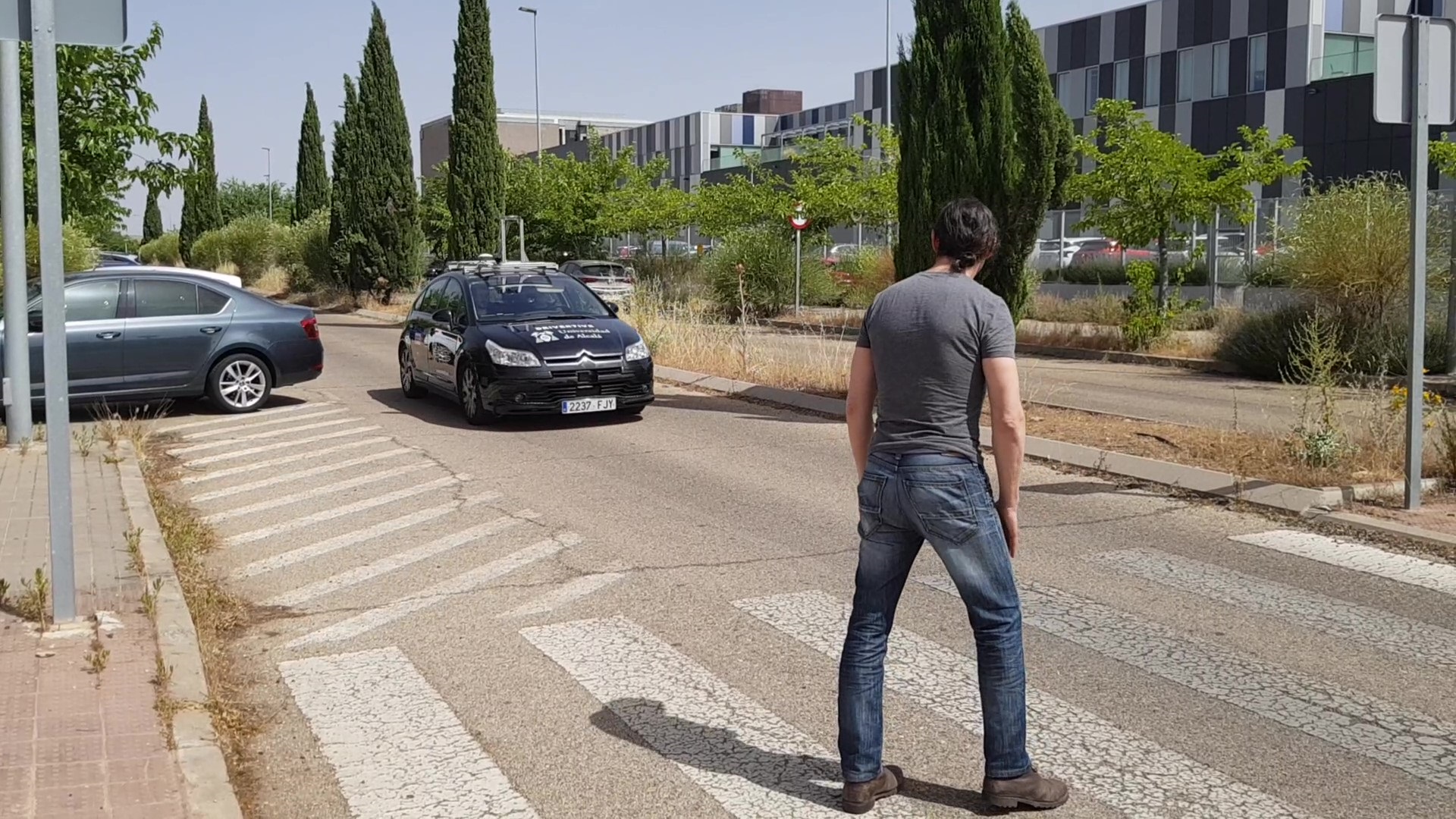}%
\label{fig_ped_6}}
\hfil
\subfloat[The Pedestrian decides to cross when the vehicle is almost stopped.]{\includegraphics[width=0.24\textwidth]{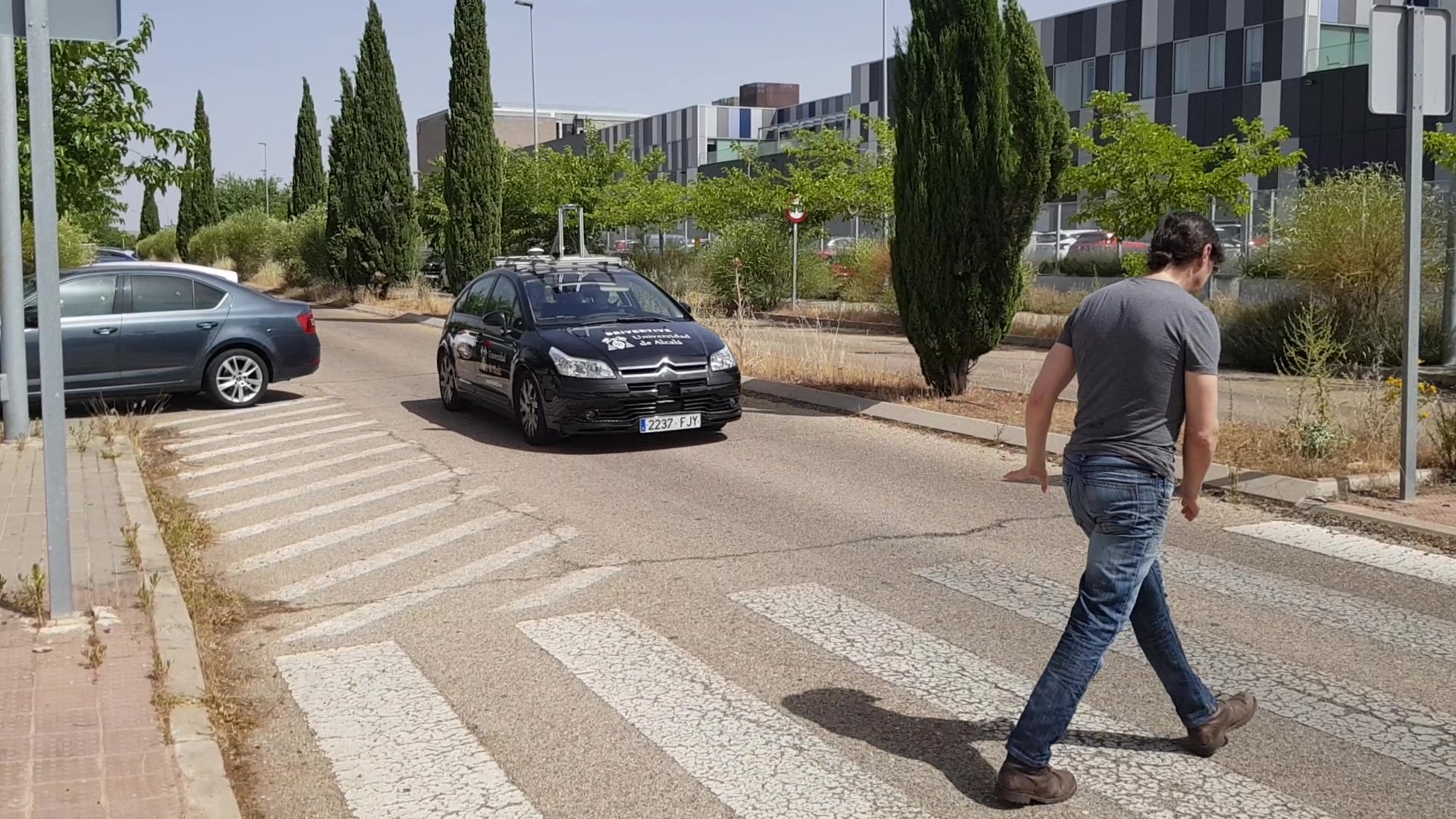}%
\label{fig_ped_7}}
\hfil
\subfloat[The Pedestrian crosses the crosswalk.]{\includegraphics[width=0.24\textwidth]{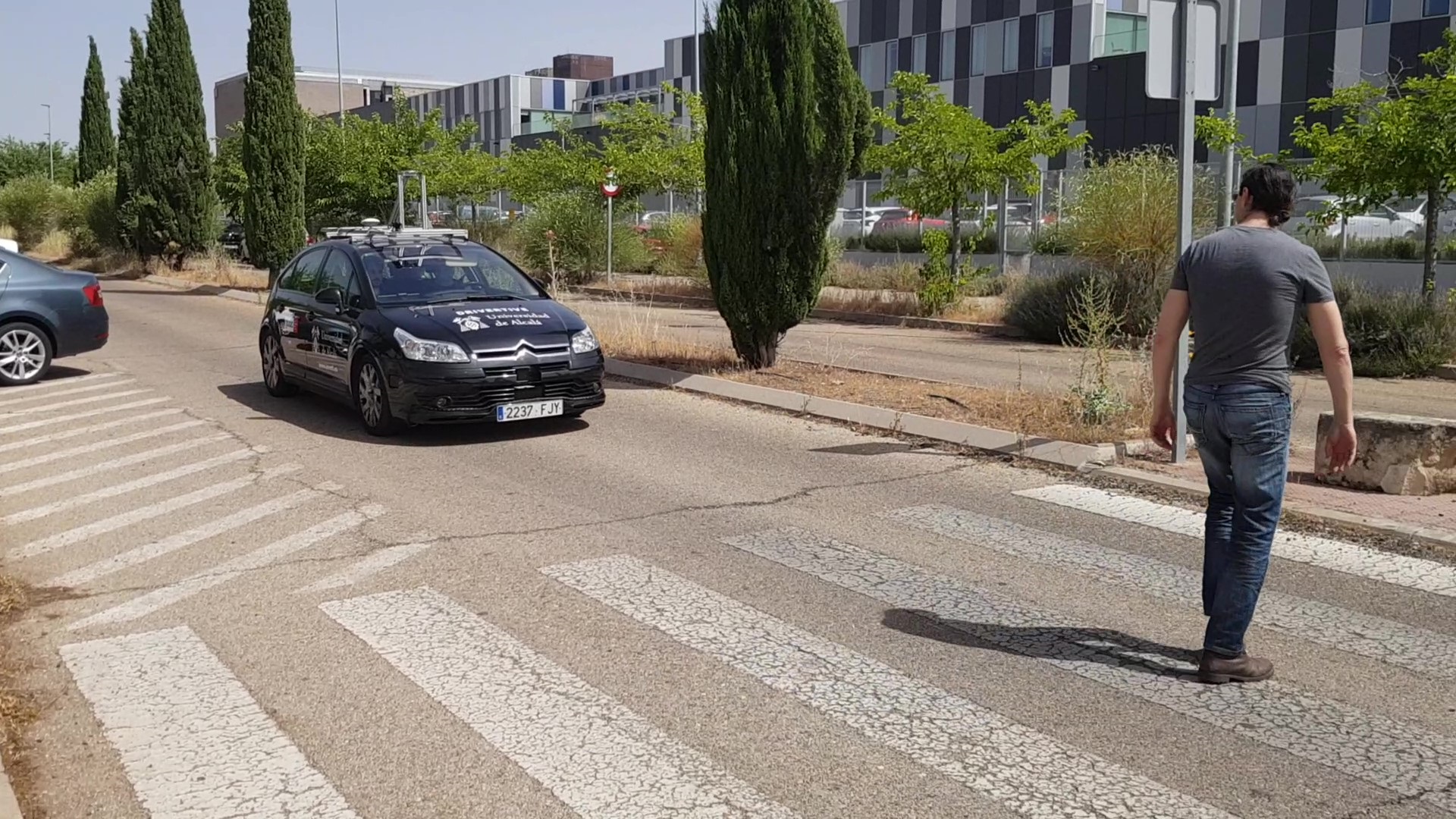}%
\label{fig_ped_8}}
\caption{Example of vehicle-pedestrian interaction - exterior camera.}
\label{fig:example_interacion_exterior}
\end{figure*}

The crossing decision event is defined as the moment in which the pedestrian makes the decision to cross. 
We follow the hypothesis that the decision to cross is a hidden state that has an external and a posterior manifestation that can be observed. The “delay” between the decision and the external manifestation can vary depending on the person and the situation. Alternatively to the crossing decision event, we propose to use the crossing event as the metric to evaluate the behavior of the pedestrian using direct measurements in the experiments. The crossing event is defined as the frame in which the pedestrian enters the vehicle lane and physically exposes his/her body to a potential and real injury. The background idea is that an early crossing decision will produce an early crossing event and a late crossing decision will produce a late crossing event. The main difference is that the crossing event is not a hidden state. It is directly observable and can be unequivocally identified when the vehicle lane is defined using the lane markings. Figure \ref{fig:vehicle_lane} shows the vehicle lane boundary and the crossing event, defined as the moment in which the pedestrian enters the vehicle’s lane. See figure \ref{fig:example_interacion_exterior} for a crossing sequence description example. 

%%%%%%%%%%%%%%%%%%%%%%%%%%%%%%%%%%%%%%%%%%%%%%%%%%%%%%%%%%%%%%%%%%%%%%%%%%%%%%%
%%%%%%%%%%%%%%%%%%%%%%%%%%%%%%%%%%%%%%%%%%%%%%%%%%%%%%%%%%%%%%%%%%%%%%%%%%%%%%%

\section{Results}
This section presents and analyzes the responses to the questionnaire and the measured variables for each type of interaction with the vehicle. For the analysis of the questionnaires, descriptive statistics and the Wilcoxon signed-rank test has been used. For the analysis of the direct measures, the Student t-test has been used to extract information.

\subsection{Questionnaire Results}
Questions from Q1 to Q6, are analyzed using the Wilcoxon Signed Rank test. Question 7 is a choice answer and its analysis is limited to the mode values. 

The analysis of the participants’ answers to questions from Q1 to Q6 will follow the same structure for each question. The alternative hypothesis matrix has been selected as the way to express categorical statements comparing the answers obtained after each experiment. As the null hypothesis, we propose $H_0:\mu_i\leq\mu_j$ and as the alternative hypothesis, we take $H_1:\mu_i>\mu_j$. 
A check-mark in a specific cell in table \ref{tab:wilcoxon} means that $H_0$ is rejected and $H_1$ is accepted when comparing the answers provided in test $i$ (left column) with test $j$ (top row). In this specific context, the rejection of $H_0$ means that there is a difference with statistical significance between the answers for test $i$ and $j$, and the answers for test $i$ have a higher score in the Likert scale than for test $j$. The significance level has been selected to $\alpha$=0.05 for all the probes. Experiment numbers are defined in table \ref{tab:testsetup}.

\begin{table}[htbp]
\renewcommand{\arraystretch}{1.1}
\caption{Wilcoxon Signed Rank test for Questions Q1-Q6}
\begin{center}
\begin{tabular}{ccc|p{1cm}p{1cm}p{1cm}p{1cm}}
%\hline
\multicolumn{3}{c|}{\textbf{$H_1:\mu_i>\mu_j$}}& \multicolumn{4}{c}{\textbf{Test number $j$}} \\
%\cline{3-7} 
 \multicolumn{3}{c|}{} & 1 & 2 & 3 & 4\\
\hline
% & & %\multicolumn{4}{|c|}{\textbf{Question Q1}} \\
%\cline{3-7}
\multirow{24}{*}{\rotatebox[origin=c]{90}{\textbf{Test number $i$}}} & \multirow{4}{*}{\rotatebox[origin=c]{90}{\textbf{Q1}}}     & 1   & --   & \checkmark    &       & \\
                                            &&2   &     & --    &      & \\
                                            &&3   & \checkmark   & \checkmark     & --    & \checkmark\\
                                            &&4   &     & \checkmark     &      & --\\

\cline{3-7}
&\multirow{4}{*}{\rotatebox[origin=c]{90}{\textbf{Q2}}} & 1   & --   &      &      & \\
                                                        &&2   & \checkmark    & --    & \checkmark     & \\
                                                        &&3   &     &      & --    & \\
                                                        &&4   & \checkmark    &      & \checkmark     & --    \\

\cline{3-7}
&\multirow{4}{*}{\rotatebox[origin=c]{90}{\textbf{Q3}}} & 1   & --   &      &      & \\
                                                        &&2   &     & --    &      &      \\
                                                        &&3   & \checkmark    & \checkmark     & --    & \checkmark     \\
                                                        &&4   & \checkmark    & \checkmark     &      & --    \\

\cline{3-7}
&\multirow{4}{*}{\rotatebox[origin=c]{90}{\textbf{Q4}}} & 1   & --   & \checkmark     &      & \checkmark\\
                                                        &&2   &     & --    &      &      \\
                                                        &&3   &     & \checkmark     & --    & \checkmark     \\
                                                        &&4   &     & \checkmark     &      & --    \\

\cline{3-7}
&\multirow{4}{*}{\rotatebox[origin=c]{90}{\textbf{Q5}}} & 1   & --   &      &      & \\
                                                        &&2   & \checkmark    & --    & \checkmark     &      \\
                                                        &&3   & \checkmark    &      & --    &      \\
                                                        &&4   & \checkmark    &      & \checkmark     & --    \\

\cline{3-7}
&\multirow{4}{*}{\rotatebox[origin=c]{90}{\textbf{Q6}}} & 1   & --   &      &      & \\
                                                        &&2   &     & --    &      &      \\
                                                        &&3   & \checkmark    & \checkmark     & --    &      \\
                                                        &&4   & \checkmark    & \checkmark     &      & --    \\
\end{tabular}
\label{tab:wilcoxon}
\end{center}
\end{table}

Table \ref{tab:q7_freq} summarizes as a frequency table the answers for question Q7. It can be observed that the preferred communication mode for the internal HMI is the combination of audio and video simultaneously.

\begin{table}[htbp]
\renewcommand{\arraystretch}{1.1}
\caption{Frequency Table for Question 7 Answers}
\begin{center}
\begin{tabular}{c|p{0.9cm}p{0.9cm}p{0.9cm}p{0.9cm}p{0.9cm}}
& \multicolumn{4}{c}{\textbf{Test number}}\\
\textbf{Answer} & 0 & 1 & 2 & 3 & 4 \\
\hline
Nothing         & 32 & 32    & 32 & 1 & 1 \\
Audio           & 0    & 0     & 0  & 7 & 10 \\
Video           & 0    & 0     & 0  & 9 & 9 \\
Both            & 0    & 0     & 0  & 15& 12 \\
\end{tabular}
\label{tab:q7_freq}
\end{center}
\end{table}

\subsection{Direct Measurements Results}
The analysis of the direct measures has been performed following the same alternative hypothesis matrix and the Student-t test. All the experiment measures are compared with all of them to obtain in which of them statistical significance can be observed. All the variables are continuous, consequently, the student’s t-test for paired samples has been used. The same null $H_0:\mu_i\leq\mu_j$ and alternative $H_1:\mu_i>\mu_j$ hypothesis are proposed for the alternative hypothesis matrix. A-check mark a specific cell in table \ref{tab:student} means that $H_0$ is rejected and $H_1$ when comparing the measures from test $i$ (left columns) against $j$ (top row). In this specific context, a check-mark means that the distance, speed or TTC in test $i$ is higher than for test $j$ with a significance level $\alpha=0.05$. To remove possible outliers and errors from different sources the two highest and lowest values have been removed from the analysis for all the variables.

\begin{table}[htbp]
\renewcommand{\arraystretch}{1.1}
\caption{Student t-test for Distance, Speed, and TTC at the Crossing Event}
\begin{center}
\begin{tabular}{ccc|p{1cm}p{1cm}p{1cm}p{1cm}}
%\hline
\multicolumn{3}{c|}{\textbf{$H_1:\mu_i>\mu_j$}}& \multicolumn{4}{c}{\textbf{Test number $j$}} \\
%\cline{3-7} 
 \multicolumn{3}{c|}{} & 1 & 2 & 3 & 4\\
\hline
% & & %\multicolumn{4}{|c|}{\textbf{Question Q1}} \\
%\cline{3-7}
\multirow{12}{*}{\rotatebox[origin=c]{90}{\textbf{Test number $i$}}} & \multirow{4}{*}{\rotatebox[origin=c]{90}{\textbf{Distance}}}   & 1   & --   & \checkmark    &       & \checkmark\\
                                                &&2   &     & --    &      & \\
                                                &&3   & \checkmark   & \checkmark     & --    & \checkmark\\
                                                &&4   &     &      &      & --\\
\cline{3-7}
&\multirow{4}{*}{\rotatebox[origin=c]{90}{\textbf{Speed}}}  & 1   & --   & \checkmark     &      & \checkmark\\
                                                            &&2   &     & --    &      &      \\
                                                            &&3   & \checkmark    & \checkmark     & --    & \checkmark     \\
                                                            &&4   &     &      &      & --    \\
\cline{3-7}
&\multirow{4}{*}{\rotatebox[origin=c]{90}{\textbf{TTC}}}    & 1   & --   & \checkmark     &      & \checkmark\\
                                                            &&2   &     & --    &      & \checkmark     \\
                                                            &&3   & \checkmark    & \checkmark     & --    & \checkmark     \\
                                                            &&4   &     &      &      & --    \\
\end{tabular}
\label{tab:student}
\end{center}
\end{table}

\begin{figure}
\centerline{\includegraphics[width=\columnwidth]{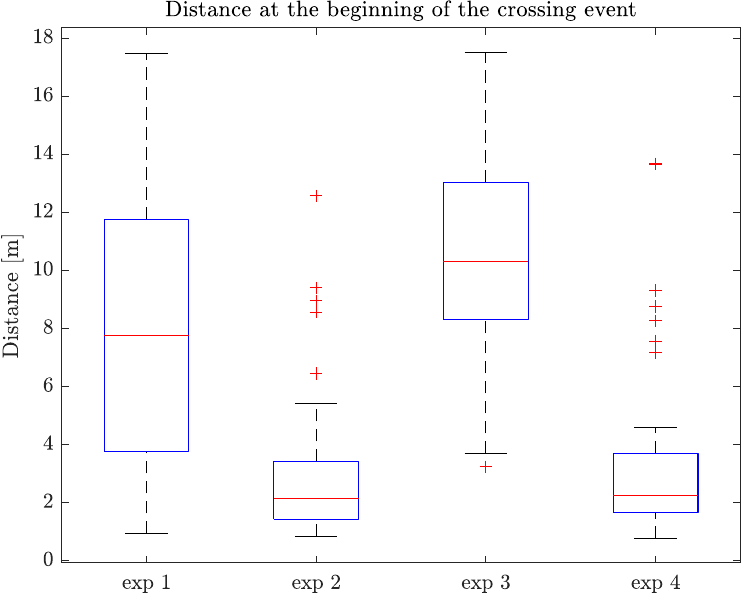}}
\caption{Boxplot for distance variable at the crossing event.}
\label{fig:boxplot_distance}
\end{figure}

\vspace{-0.1cm}
\subsection{Results Discussion}
\vspace{-0.1cm}

Based on the results shown in table \ref{tab:wilcoxon} we can make the following statements:
The optimal braking maneuver contributes to increasing the pedestrian's confidence in the vehicle (Q1: t1 vs t2 and t3 vs t4).
The eHMI contributes to increasing the pedestrian's confidence in the vehicle. (Q1: t3 vs t1 and t4 vs t2)
Pedestrians perceived the aggressive braking maneuvers as “more aggressive” or “less conservative” than the gentle breaking maneuvers (Q2: t2 vs t1 and t4 vs t3).
The optimal braking maneuver contributes to increasing the passenger's confidence in the vehicle (Q4: t1 vs t2 and t3 vs t4).
The iHMI contributes to increasing the passenger's confidence in the vehicle when stopping in a crosswalk with an aggressive braking maneuver (Q4: t4 vs t2).
Passengers perceived the aggressive braking maneuvers as “more aggressive” or “less conservative” than the gentle braking maneuvers (Q5: t2 vs t1 and t4 vs t3).
We cannot state that the iHMI contributes to increasing the passenger's confidence in the vehicle when stopping in a crosswalk with an optimal braking maneuver (Q4: t3 vs t1).
Attending to the frequency values provided for Q7 we can state that passengers found the combined mode (audio plus video) most helpful to build confidence than the audio or video choices for the iHMI.
Based on the results shown in table \ref{tab:student} we can make the following statements:
The optimal braking maneuver contributes to increasing the distance at the crossing event when crossing in the crosswalk(dist.: t1 vs t2 and t3 vs t4).
The eHMI in combination with the optimal braking maneuver contributes to increasing the distance at the crossing event (dist.: t3 vs t1).
We cannot state that the eHMI contributes to increasing the distance at the crossing event when the AV brakes with the aggressive braking maneuver (dist.: t4 vs t2). This suggests that subjectively pedestrians feel safer but objectively they are not and the perception of risk dominates their actual behavior. As an additional finding, there were no statistical differences between the responses and measures obtained from the participants regardless of their initial role played.

%The results for the alternative hypothesis matrix based on speed or TTC at the crossing event confirm the same statements that have been already made based on distance.

%%%%%%%%%%%%%%%%%%%%%%%%%%%%%%%%%%%%%%%%%%%%%%%%%%%%%%%%%%%%%%%%%%%%%%%%%%%%%%%
%%%%%%%%%%%%%%%%%%%%%%%%%%%%%%%%%%%%%%%%%%%%%%%%%%%%%%%%%%%%%%%%%%%%%%%%%%%%%%%

\section{Conclusions and Future Work}
%This section provides the conclusions of the study based on the results and its analysis. 
\vspace{-0.1cm}
Questionnaires and direct measurements have proven that the iHMI and eHMI in combination with "gentle" or early braking maneuvers help to build trust in the AV when interacting with a pedestrian in a crosswalk for both the pedestrian and the passenger. However, there is a relevant difference between conclusions derived from questionnaires and measured variables. When we compare the experiments with the aggressive braking maneuver pedestrians express more confidence when using the eHMI than when not using it. However, it does not result in an earlier crossing event and consequently in an earlier crossing decision.

As a future work, this study can be extended by including interactions in non-signalized crossing areas and can be also replicated in Virtual Reality (VR) to study the gap between real and virtual interactions.

%%%%%%%%%%%%%%%%%%%%%%%%%%%%%%%%%%%%%%%%%%%%%%%%%%%%%%%%%%%%%%%%%%%%%%%%%%%%%%%
%%%%%%%%%%%%%%%%%%%%%%%%%%%%%%%%%%%%%%%%%%%%%%%%%%%%%%%%%%%%%%%%%%%%%%%%%%%%%%%

\section*{Acknowledgment}
\vspace{-1mm}
This work was funded by the HUMAINT project, Directorate-General
Joint Research Centre of the European Commission and by Research Grants PID2020-114924RB-I00 and PDC2021-121324-I00 (Spanish Ministry of Science and
Innovation) and S2018/EMT-4362 SEGVAUTO
4.0-CM and CM/JIN/2021-005 (Community of Madrid).

\section*{Disclaimer}
\vspace{-1mm}
The views expressed in this article are purely those of the authors and may not, under any circumstances, be regarded as 
an official position of the European Commission. 

\bibliographystyle{IEEEtran}
\bibliography{citations.bib}

\begin{thebibliography}{10}
\providecommand{\url}[1]{#1}
\csname url@rmstyle\endcsname
\providecommand{\newblock}{\relax}
\providecommand{\bibinfo}[2]{#2}
\providecommand\BIBentrySTDinterwordspacing{\spaceskip=0pt\relax}
\providecommand\BIBentryALTinterwordstretchfactor{4}
\providecommand\BIBentryALTinterwordspacing{\spaceskip=\fontdimen2\font plus
\BIBentryALTinterwordstretchfactor\fontdimen3\font minus
  \fontdimen4\font\relax}
\providecommand\BIBforeignlanguage[2]{{%
\expandafter\ifx\csname l@#1\endcsname\relax
\typeout{** WARNING: IEEEtran.bst: No hyphenation pattern has been}%
\typeout{** loaded for the language `#1'. Using the pattern for}%
\typeout{** the default language instead.}%
\else
\language=\csname l@#1\endcsname
\fi
#2}}

\bibitem{Li2019}
M.~Li, B.~E. Holthausen, R.~E. Stuck, and B.~N. Walker, ``No risk no trust:
  Investigating perceived risk in highly automated driving,'' in
  \emph{Proceedings of the 11th AutomotiveUI Conference}.\hskip 1em plus 0.5em
  minus 0.4em\relax Association for Computing Machinery, 2019, p. 177–185.

\bibitem{Llorca2021}
D.~Fernández~Llorca and E.~Gómez, ``Trustworthy autonomous vehicles,''
  \emph{EUR 30942 EN, Publications Office of the European Union, Luxembourg},
  vol. JRC127051, 2021.

\bibitem{Detjen2021}
H.~Detjen, S.~Faltaous, B.~Pfleging, S.~Geisler, and S.~Schneegass, ``How to
  increase automated vehicles' acceptance through in-vehicle interaction
  design: A review,'' \emph{International Journal of Human-Computer
  Interaction}, vol.~37, no.~4, pp. 308--330, 2021.

\bibitem{Rasouli2020}
A.~{Rasouli} and J.~K. {Tsotsos}, ``Autonomous vehicles that interact with
  pedestrians: A survey of theory and practice,'' \emph{IEEE Transactions on
  Intelligent Transportation Systems}, vol.~21, no.~3, pp. 900--918, 2020.

\bibitem{ignacio_v2v}
I.~Parra, H.~Corrales, N.~Hern{\'a}ndez, S.~Vigre, D.~Llorca, and M.~A. Sotelo,
  ``Performance analysis of vehicle-to-vehicle communications for critical
  tasks in autonomous driving,'' in \emph{2019 IEEE Intelligent Transportation
  Systems Conference (ITSC)}, 2019, pp. 195--200.

\bibitem{Dey2021}
A.~{Rasouli} and J.~K. {Tsotsos}, ``Autonomous vehicles that interact with
  pedestrians: A survey of theory and practice,'' \emph{IEEE Transactions on
  Intelligent Transportation Systems}, vol.~21, no.~3, pp. 900--918, 2020.

\bibitem{Martin2023}
S.~Martín, R.~Izquierdo, I.~García~Daza, M.~A. Sotelo, and
  D.~Fernández~Llorca, ``Digital twin in virtual reality for human-vehicle
  interactions in the context of autonomous driving,'' in \emph{IEEE
  Intelligent Transportation Systems Conference}, 2023.

\bibitem{Lagstrom2015}
T.~Lagstrom and V.~M. Lundgren, \emph{AVIP-autonomous vehicles interaction with
  pedestrians}.\hskip 1em plus 0.5em minus 0.4em\relax Gothenborg, Sweden: M.S.
  thesis, Dept. Product-Prod. Develop., 2015.

\bibitem{drivertive2018}
I.~Parra, R.~Izquierdo, J.~Alonso, A.~García-Morcillo, D.~Fernández-Llorca,
  and M.~A. Sotelo, ``The experience of drivertive-driverless cooperative
  vehicle-team in the 2016 gcdc,'' \emph{IEEE Transactions on Intelligent
  Transportation Systems}, vol.~19, no.~4, pp. 1322--1334.

\bibitem{dijksterhuis2015impact}
C.~Dijksterhuis, B.~Lewis-Evans, B.~Jelijs, D.~de~Waard, K.~Brookhuis, and
  O.~Tucha, ``The impact of immediate or delayed feedback on driving behaviour
  in a simulated pay-as-you-drive system,'' \emph{Accident Analysis \&
  Prevention}, vol.~75, pp. 93--104, 2015.

\bibitem{tom2011gender}
A.~Tom and M.-A. Grani{\'e}, ``Gender differences in pedestrian rule compliance
  and visual search at signalized and unsignalized crossroads,'' \emph{Accident
  Analysis \& Prevention}, vol.~43, no.~5, pp. 1794--1801, 2011.

\bibitem{bazilinskyy2019survey}
P.~Bazilinskyy, D.~Dodou, and J.~De~Winter, ``Survey on ehmi concepts: The
  effect of text, color, and perspective,'' \emph{Transportation research part
  F: traffic psychology and behaviour}, vol.~67, pp. 175--194, 2019.

\bibitem{chang2022can}
C.-M. Chang, K.~Toda, X.~Gui, S.~H. Seo, and T.~Igarashi, ``Can eyes on a car
  reduce traffic accidents?'' in \emph{Proceedings of the 14th international
  conference on automotive user interfaces and interactive vehicular
  applications}, 2022, pp. 349--359.

\bibitem{modes5}
D.~Moore, R.~Currano, G.~E. Strack, and D.~Sirkin, ``The case for implicit
  external human-machine interfaces for autonomous vehicles,'' in
  \emph{Proceedings of the 11th international conference on automotive user
  interfaces and interactive vehicular applications}, 2019, pp. 295--307.

\bibitem{mason2022lighting}
B.~Mason, S.~Lakshmanan, P.~McAuslan, M.~Waung, and B.~Jia, ``Lighting a path
  for autonomous vehicle communication: the effect of light projection on the
  detection of reversing vehicles by older adult pedestrians,''
  \emph{International journal of environmental research and public health},
  vol.~19, no.~22, p. 14700, 2022.

\bibitem{lynch2016social}
M.~Lynch, ``Social constructivism in science and technology studies,''
  \emph{Human Studies}, vol.~39, pp. 101--112, 2016.

\bibitem{benderius2017best}
O.~Benderius, C.~Berger, and V.~M. Lundgren, ``The best rated human--machine
  interface design for autonomous vehicles in the 2016 grand cooperative
  driving challenge,'' \emph{IEEE Transactions on intelligent transportation
  systems}, vol.~19, no.~4, pp. 1302--1307, 2017.

\bibitem{murali2022intelligent}
P.~K. Murali, M.~Kaboli, and R.~Dahiya, ``Intelligent in-vehicle interaction
  technologies,'' \emph{Advanced Intelligent Systems}, vol.~4, no.~2, p.
  2100122, 2022.

\bibitem{he2017mask}
K.~He, G.~Gkioxari, P.~Doll{\'a}r, and R.~Girshick, ``Mask r-cnn,'' in
  \emph{Proceedings of the IEEE international conference on computer vision},
  2017, pp. 2961--2969.

\bibitem{GRAIL}
R.~Izquierdo, C.~Salinas, J.~Alonso, I.~Parra, D.~Fern\'{a}ndez-Llorca, and
  M.~A. Sotelo, ``Testing predictive automated driving systems: Lessons learned
  and future recommendations,'' \emph{IEEE Intelligent Transportation Systems
  Magazine}, vol.~14, no.~6, pp. 77--93, 2022.

\bibitem{joshi2015likert}
A.~Joshi, S.~Kale, S.~Chandel, and D.~K. Pal, ``Likert scale: Explored and
  explained,'' \emph{British journal of applied science \& technology}, vol.~7,
  no.~4, p. 396, 2015.

\end{thebibliography}
% \begin{thebibliography}{00}
% \bibitem{b1} G. Eason, B. Noble, and I. N. Sneddon, ``On certain integrals of Lipschitz-Hankel type involving products of Bessel functions,'' Phil. Trans. Roy. Soc. London, vol. A247, pp. 529--551, April 1955.
% \bibitem{b2} J. Clerk Maxwell, A Treatise on Electricity and Magnetism, 3rd ed., vol. 2. Oxford: Clarendon, 1892, pp.68--73.
% \bibitem{b3} I. S. Jacobs and C. P. Bean, ``Fine particles, thin films and exchange anisotropy,'' in Magnetism, vol. III, G. T. Rado and H. Suhl, Eds. New York: Academic, 1963, pp. 271--350.
% \bibitem{b4} K. Elissa, ``Title of paper if known,'' unpublished.
% \bibitem{b5} R. Nicole, ``Title of paper with only first word capitalized,'' J. Name Stand. Abbrev., in press.
% \bibitem{b6} Y. Yorozu, M. Hirano, K. Oka, and Y. Tagawa, ``Electron spectroscopy studies on magneto-optical media and plastic substrate interface,'' IEEE Transl. J. Magn. Japan, vol. 2, pp. 740--741, August 1987 [Digests 9th Annual Conf. Magnetics Japan, p. 301, 1982].
% \bibitem{b7} M. Young, The Technical Writer's Handbook. Mill Valley, CA: University Science, 1989.
% \end{thebibliography}

\end{document}